# Large excitonic binding energy in GaN based superluminescent light emitting diode on naturally survived sub-10 nm lateral nanowires

**Debashree Banerjee[1,2], Maharaja B. Nadar[1], Pankaj Upadhyay[1], Raksha Singla[1], Sandeep Sankaranarayanan[1], Dolar Khachariya[1], Nakul Pande[1], Kuldeep Takhar[1], Swaroop Ganguly[1,2] and Dipankar Saha[1,2,*]**

[1] *Applied Quantum Mechanics Laboratory, Center of Excellence in Nanoelectronics, Department of Electrical Engineering, Indian Institute of Technology Bombay, Mumbai 400076, India*

[2] *IITB-Monash Research Academy, Indian Institute of Technology Bombay, Mumbai 400076, India*

[*]*dipankarsaha@iitb.ac.in*

**Abstract:** We demonstrate a novel method for nanowire formation by natural selection during wet chemical etching in boiling Phosphoric acid. It is observed that wire lateral dimensions of sub-10 nm and lengths of 700 nm or more have been naturally formed during the wet etching. The dimension variation is controlled through etching times wherein the underlying cause is the merging of the nearby crystallographic hexagonal etch pits. The emission processes involving excitons are found to be efficient and lead to enhanced emission characteristics. The exciton binding energy is augmented by using quantum confinement whereby enforcing greater overlap of the electron-hole wave-function. The surviving nanowires are nearly defect-free, have large exciton binding energies of around 45 meV and a small temperature variation of the output electroluminescent light. We have observed superluminescent behaviour of the LEDs formed on these nanowires. There is no observable efficiency roll off till current densities of 400 A/cm$^2$. The present work thus provides an innovative and cost effective manner of device fabrication on the formed nanowires and proves the immediate performance enhancement achievable.

**1. Introduction**

Radiative recombination processes in semiconductors that involve excitons give rise to narrow emission peaks and greater efficiencies even at room temperature, provided that the excitons survive till such high temperatures. Excitons are the bound states of the electron-hole (e-h) pair due to the unscreened Coulomb attraction between the oppositely charged carriers. The excitons are formed when the e-h pair lowers its energy by an amount equal to the exciton binding energy. The larger the exciton binding energy the greater is the stability of the exciton against thermal dissociation. The wide bandgap semiconductors like Gallium Nitride (GaN) and Zinc Oxide (ZnO) have large exciton binding energies ( > 26 meV), which permit their existence at room temperature. Indium Gallium Nitride (InGaN), the ternary alloy of Indium Nitride (InN) and GaN, have been successfully utilized in the fabrication of light emitting diodes (LEDs) and lasers [1-4]. Localized excitons engendered by the Indium clusters in InGaN make it possible to extract reasonably efficient light emission from the InGaN-based opto-electronic devices in spite of the presence of a high defect density in the material. These pseudo-confinement regions capture the excitons and prevent their movement towards the defects. Thus the optical efficiencies presently observed in the LEDs and lasers can be enhanced manifold if the InGaN device active region is rid of the structural defects, since they promote non-radiative recombination. The presence of polarization fields in GaN tend to cause spatial separation of the e-h pairs and inhibit exciton formation. The exciton binding energy can be augmented by using quantum confinement whereby enforcing greater overlap of the electron-hole wave-function and strain relaxation due to enhanced surface area. It has been demonstrated experimentally that InGaN-GaN quantum wells of dimensions smaller than the exciton Bohr radius (3.4 nm in InGaN) exhibit highly reduced piezoelectric fields while allowing pronounced exciton localization effects [5]. Efficiency roll-off at high current densities in GaN-based LEDs is a technology challenge that is yet to be comprehensively addressed. Plasmon couplings, use of m-plane non-polar growth directions, ensuring low defect densities in the active region with good lattice match between the wells and barriers and mitigating carrier leakage are some of the possible ways to alleviate the problem of droop in GaN LEDs [6-8].

It has been observed that wet chemical etching in boiling Phosphoric acid ($H_3PO_4$) selectively removes the threading dislocations in GaN resulting in hexagonal etch pits for Ga-polar samples [9]. The remaining material is almost entirely free of these defects. Among all other defects threading dislocations are one of the major contributors of non-radiative recombination. Wet etching thus provides a method of obtaining high quality materials. Additionally, the formation of the etch pits have been shown to increase the light output efficiency of the LEDs probably due to increase in light extraction efficiency [10]. Another approach to overcome efficiency droop is to use the optical gain in a material to enhance the light output in the LED while not allowing it to lase by eliminating the feedback mechanism – the path of the superluminescent LEDs (SLEDs) [11]. Feedback implies a mechanism to preferentially allow one or some of the many different wavelengths' radiation emitted by the active region to grow stronger in intensity while the non-favoured ones die out subsequently. By eliminating the necessity to match the resonant criterion, all modes of the emitted light survive to give a wide-band optical output. The noise features of LEDs are absent in SLEDs while it offers better coupling to optical fibers [12].

The quantum confined features which are attractive for opto-electronic devices can be obtained in two ways: epitaxial growth or patterned dry etching. Vapour-Liquid-Solid (VLS) or Stranski–Krastanov growth methods are used in the bottom-up approach, while dry etching of the patterned heterostructure yield wires and dots in the top-down scheme. Here we demonstrate a novel method for nanowire formation by the use of the crystallographic wet chemical etching process that was previously used for surface texturing of the conventional GaN LEDs [10, 13, 14] or for smoothening the surface after dry etching to reduce the surface damage [15].The low density of the nanowires is further advantageous to study quantum mechanical phenomena, which becomes complicated if there is a forest of nanowires when grown in a bottom-up approach. Vertical nanowire LEDs have been used for alleviation of the droop in GaN-based devices [16], however in the present report, the external quantum efficiency and light output power versus bias current density of the lateral single-nanowire LED has no droop.

In the present report we demonstrate the formation of lateral nanowires of widths less than 10 nm in an InGaN-based quantum well LED heterostructure by wet chemical etching in boiling Phosphoric acid. The hexagonal etch pits form at threading dislocation sites where the anisotropic wet etching commence, with longer etch times the etch pits widen and adjacent pits nearly merge. The thin wall of material that remains between two nearby hexagonal etch pits can be constricted to widths of practically any dimension by controlling the etching time. These narrow and defect free regions of the quantum well LED heterostructure can be ideal candidates for making highly efficient superluminescent LEDs and lasers. Photoluminescence (PL) studies show an increase of the exciton binding energies from 20.8 meV in the quantum well sample to 45.9 meV in the quantum wire sample. This corroborates the fact that the wire lateral width decreases with etch time since the decreasing wire width leads to tighter coupling of the carrier wave-functions that strengthens the exciton binding. An edge emitting LED is fabricated with the active region being the single lateral nanowire having a vertical quantum well structure. The top n-doped layer is kept intentionally thicker to avoid complete removal of the n-region in the n-i-p LED heterostructure during the blanket wet etching. The light output power versus bias current (L-I) characteristics at room temperature show no droop for higher current densities, for lower current densities the light output is small and it increases super-linearly beyond a threshold current density of ~ 140 A/cm$^2$ while the electroluminescent (EL) spectrum is broadband. We have observed that the relative blue shift of the primary energy peak of the EL spectrum from the nanowire LED (NW-LED) compared to the quantum well control LED is only 10 meV. The blue shift value is rather small considering that the wire lateral width is extremely narrow. The apparent anomaly can be explained with an infinite quantum well model coupled to the condition that the active region optical emission involves the lowering of the radiative emission energy due to the binding energy of the exciton.

## 2. Experimental Results

The LED heterostructure schematic in Fig. 1(a) is a p-i-n structure with the active region consisting of the six $In_xGa_{1-x}N$ (10 nm)/$In_yGa_{1-y}N$ (3 nm), with x=0.01 and y=0.1, intrinsic quantum wells grown on c-plane sapphire substrate by the metalorganic chemical vapour deposition (MOCVD) technique. A 25 nm GaN buffer layer is grown on the sapphire substrate followed by a 1.5 µm thick undoped GaN, a 2 µm Mg-doped p-GaN (carrier concentration ~ 1.2 x $10^{20}$ cm$^{-3}$) and a 10 nm p-$Al_{0.12}Ga_{0.88}$N electron blocking layer (8 x $10^{19}$ cm$^{-3}$). This is followed by the active region growth and then the 100 nm thick Si-doped n-GaN (electron concentration ~ 1 x $10^{18}$ cm$^{-3}$). Phosphoric acid (concentrated $H_3PO_4$ 85 %) is slowly heated to 200° C and allowed to stabilize. The unetched quantum well sample is labelled as sample A. The InGaN-based LED hetero structures are etched in the hot reagent for 5 and 8 minutes and labelled as samples B and C. The screw type defects serve as non-radiative recombination centres but not the edge type defects [17]. It can be seen in the scanning electron microscope (SEM) image (Fig. 1(b)), that the adjacent hexagonal etch pits in sample C, identifiable as originating from screw type dislocations, has almost merged resulting in a wall that is less than 10 nm wide at the top and approximately 700 nm long (Fig. 1(c)). The wall has sloping faces and constitutes the nanowire that has the vertical quantum wells (Fig. 1(d)). Surface depletion effects are expected to reduce the wire dimension further [18]. The physical wire widths progressively diminish from sample B to C with increasing etch times. The atomic force microscopy (AFM) image in Fig. 1(e) has been used to conclude that at least 15 nm of the top n-type layer survives in sample C. The depth profile of the AFM image shows that the wire dimension at the quantum well is approximately 10 nm.

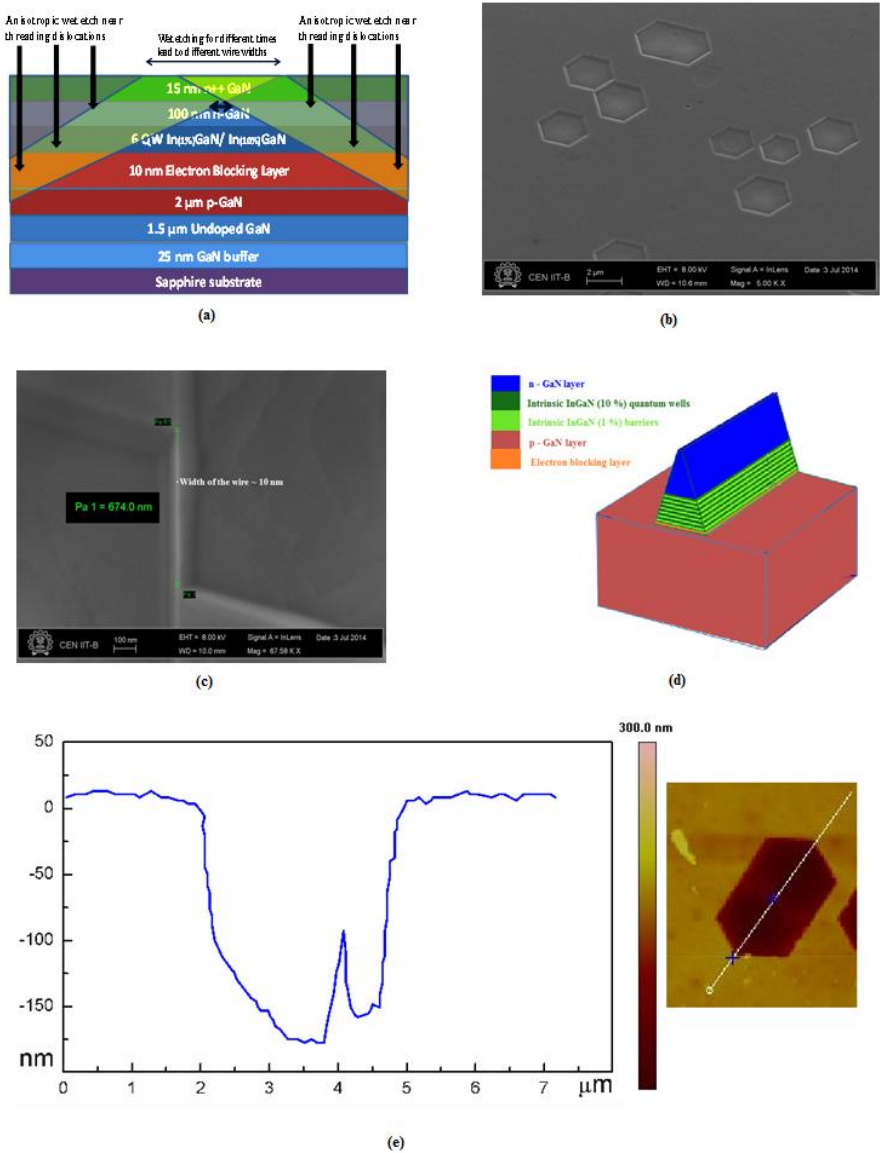

Fig. 1. (a) Schematic of the light emitting diode (LED) heterostructure and etching process and (b) the scanning electron microscope (SEM) image of the quantum wire sample after 8 minutes of etching; (c) the SEM image of a single wire whose width is approximately 10 nm and (d) The 3-dimensional schematic of the nanowire showing the sloping wall of unetched material left behind; (e) AFM image of the quantum wire sample C showing the formation of the nanowires in the regions where two adjacent etch pits have merged. The presence of dislocations provides an impetus to start the crystallographic etching where the hot $H_3PO_4$ attacks the defect and creates etch pits. As time elapses the pits widen and nearby pits nearly merge. Etching time provides a handle to control the wire width.

Temperature dependent photoluminescence (PL) studies on the samples were performed in a closed cycle Helium cryostat with the 325 nm He-Cd laser as the above band-gap excitation source. Figure 2(a) shows the low temperature PL spectrum for sample C. The broad yellow emission is probably due to Silicon doping in the n doped region of the heterostructure. The yellow emission intensity tends to saturate as the excitation power density is increased from 50.8 $W/cm^2$ to 228 $W/cm^2$, as noted from the power dependent PL spectrum at room temperature for the sample C shown in the top inset of Fig. 2(b). It is unlike the blue emission which has a monotonic increase in intensity with increasing power density of the excitation source (bottom inset of Fig. 2(b). This confirms the defect related origin of the yellow emission. The narrow line-width of the primary peak up to room temperature, the unsaturated PL intensity with increasing power density of the excitation source as well as the temperature induced blue shift indicates its excitonic genesis in sample C. As shown in Fig. 3(c), the primary peak initially shows a red shift, followed by blue shifts in the temperature range between 100 K and 200 K. It shows the temperature dependent band-gap narrowing effect above 225 K. At low temperatures when there is minimal thermal agitation, the photo-generated excitons are rendered relatively immobile in the local minima. However as the temperature rises their mobility is enhanced. Consequently, they tend to settle into the band-edge global minimas, possibly caused by Indium fluctuations, leading to the perceived red shift in the PL peak energy. However, as the temperature rises further they move out of the local energy minimas to the above bandgap quantum confined energy states and induce the anomalous blue shift of the PL peak. These localized bound states arise from the quantum confinement. The total blue shift thus indicates the presence of bound localized states and in turn is an indicator of the density of nanowires in the region of sample being probed with the laser light. For still higher temperatures the thermal bandgap narrowing takes over to give the Varshni-like temperature shift of the PL peak energy. It is observed that the magnitude of the blue shift, indicated by the energy range between the two turning points in Fig. 2(d), increases with increasing etching time which implies that the density of the localized states increase as the wires become narrower leading to the enhanced density of quantum wires on the surface. The PL primary peak energy is blue shifted in sample B and red shifted for sample C, compared to that in sample A. The exciton binding energies are obtained from the Arrhenius plot of the integrated PL intensity for all the samples [19]. It is noted that the binding energy increases from 20.8 meV for the sample A having the primary peak located at 2.82 eV to 45.9 meV in the case of sample C with the primary peak at 2.81 eV. The binding energy of excitons increases with enhanced confinement owing to better coupling between the electron-hole wave-functions in homogeneous confined structures. The nanowires are thus constituted by the vertical confining potential of the quantum well which is further confined in the lateral direction by the etching process.

The nanowires were imaged on sample C and their coordinates were ascertained before writing the mesa structures and global alignment marks on the surface using electron-beam-lithography. Subsequently the inductively coupled reactive ion etching (ICP-RIE) was done to expose the p-layer. The Ni/Au (20 nm/20 nm) Ohmic stack was deposited on the patterned sample followed by lift-off and annealing at 500° C for 1 minute in $O_2$ for the Ohmic p-contact. Following the p-Ohmic formation, local alignment marks were made on the mesa structure. The Ohmic contact to the n-layer of the nanowire consists of a narrow protrusion of and the bottom inset shows the monotonic increase of intensity of the primary peak with excitation power; (b) temperature dependent shift of the primary peak position of the PL spectrum of sample C; (c) and (d) primary peak energy versus temperature plot of the PL spectrum in Fig. 2(a). It has an S-shape indicating the presence of localized quantized states; (e) The temperature dependent light output versus excitation power density plot shows that as the temperature increases the luminescence does not decrease substantially for the primary peak. The increase of the luminescence with excitation power confirms its excitonic origin.

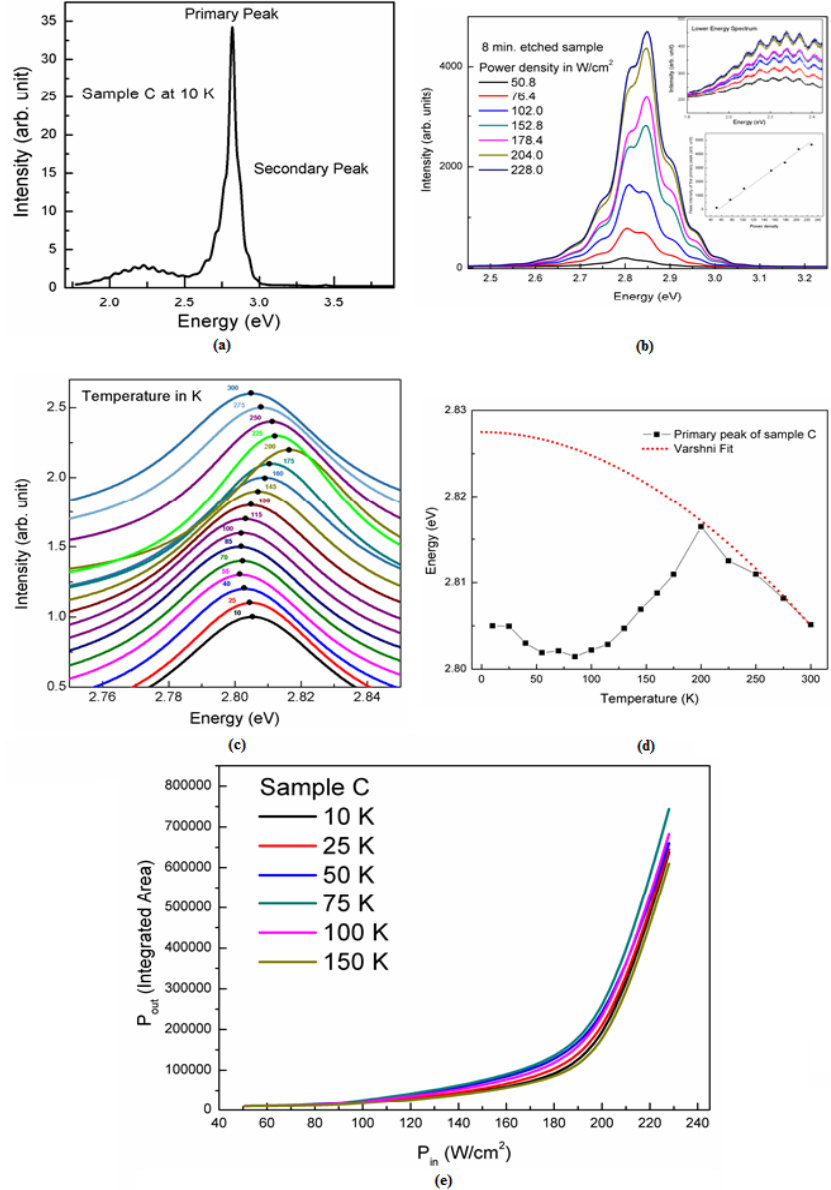

Fig. 2 Photoluminescence characteristics of an ensemble of nanowires probed by the excitation laser (a) Photoluminescence (PL) plot of the sample C (8 min. etched) at 10 K and (b) power dependent PL spectrum of sample C at 300 K, top inset to Fig. 2(b) shows the evolution of the yellow emission with excitation power density

length 4 µm and width 1 µm from a larger contact pad having a dimension of 50 µm x 50 µm. The n-Ohmic stack of Ti/Al/Ni/Au (25nm/100nm/30nm/100nm) was deposited by electron beam evaporator and annealed at $800^o$ C in $N_2$ for 90 seconds after metal lift-off. Figure 3 shows the SEM image of the nanowire LED.

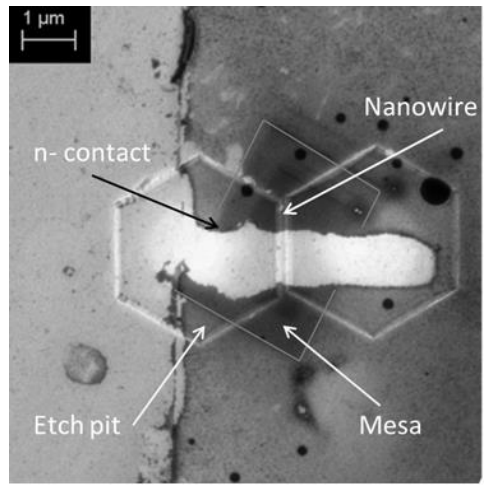

Fig. 3 SEM image of the nanowire light emitting diode. The n-contact is the Ohmic contact to the top n-region

From the temperature dependent continuous-wave plot of the light output power versus bias current density of Fig. 4(a) it is observed that the light output at lower temperature is higher than that at room temperature. However, the difference between the room temperature and 10 K L-I plots is small unlike usual GaN-based quantum well LEDs. It is intuitively understood that if the active region material is nearly free of non-radiative defects like screw dislocations then the light output will not differ much between low and room temperature operation since the major reason for poor optical performance at high temperature is the non-radiative recombination facilitated by defects. The L-I characteristic do not reveal droop up to a current density of 400 A/cm$^2$ at room temperature. The exponential nature of the L-I curve around room temperature without sharp changes in slope evince superluminescent behaviour. The intensity of the output light is probably less than the detection limits of the power-meter for low current densities in case of the nanowire LED devices. The slope efficiency of the nanowire device, defined as the slope of the experimental L-I plot, increases while that of the control LED shows a pronounced droop for similar current densities and going negative for higher current densities as can be observed from Fig. 4 (b). The bias dependent electroluminescence (EL) from the nanowire LED has an initial blue shift in the peak energy due to filling of higher energy states. Further increase of bias current density results in red shift as a result of the device heating as seen from Figure 5(c). The line width increases initially with increasing bias probably because of band filling effect as seen in Fig. 4(d). However, the FWHM drops with further increase of the bias current as expected [10]. For bias currents higher than 13.5 mA the FWHM increases again possibly due to heating effects. The observed blue shift of the NW-LED EL peak with respect to the control LED EL peak matches the theoretically obtained quantization energy for an infinitely bounded quantum well. The NW has the air dielectric on both sides as the additional confining potential wells. Hence, the NW system can be considered to be having an infinite potential well in the lateral dimension and the finite confining potential due to the vertical heterostructure. The schematic

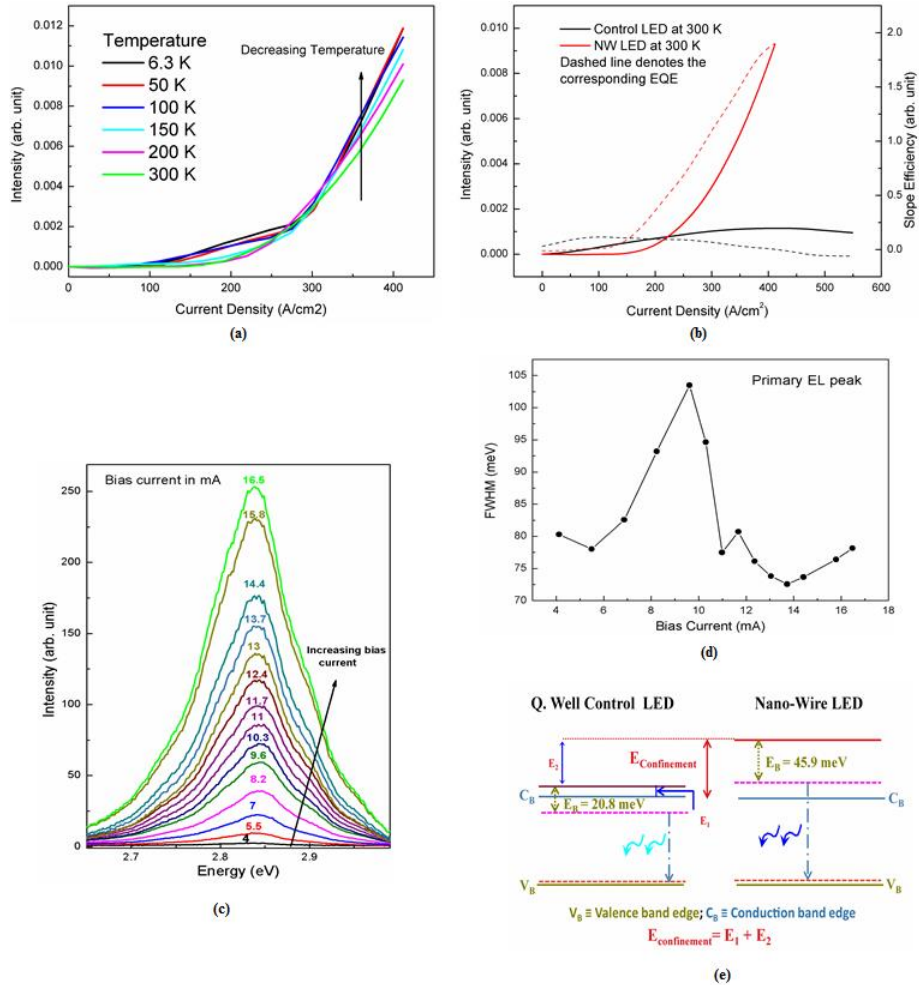

Fig. 4 (a) Temperature dependent output light power versus bias current (L-I) characteristics of the nanowire LED. Inset shows the L-I plot of the control LED at room temperature fabricated with the quantum well as-grown heterostructure and (b) room temperature output light power versus bias current density characteristics of the nanowire and control LED are denoted by solid lines while the external quantum efficiency versus bias current density is denoted by dashed lines. It is observed that the control LED has a pronounced droop; (c) bias dependent electroluminescence (EL) spectrum of the nanowire LED device at 300 K; (d) full width at half maximum (FWHM) versus bias current density of the EL primary peak and (e) schematic explaining the EL peak energy shift between the control LED and the nano-wire LED. $E_1$ is the confinement energy due to the quantum well in the LED heterostructure. The confinement energy of the nanowire LED should have led to a large blue shift of the EL energy peak i.e., $E_2$; however the binding energy of the exciton compensates a part of the blue shift. The overall observed blue shift in the EL peak is now smaller as compared to the situation where we consider only the additional confinement due to wire formation.

in Fig. 4(e) explains the reason behind the small difference in the observed excitonic emission from the wire and the control LED although the wire widths are narrow. The schematic incorporates the binding energies of 20.8 meV and 45.9 meV in the control LED and the NW LED respectively which partly compensates the expected large energy shift due to the additional confinement induced by the NW formation. The control LED gives EL peak energy of 2.84 eV, being 20.8 meV below the conduction band edge taking into account the exciton binding energy in the quantum well sample. The confinement energy for an InGaN-based NW of lateral width 8 nm is approximately 30 meV. Thus, the ground state emission energy of the NW-LED would be located at 2.89 eV. Considering the exciton binding energy of 45.9 meV associated with the primary PL peak in the quantum wire sample, we can calculate the EL emission energy to be 2.84 eV. This is in good agreement with the observed EL peak energy of 2.85 eV. The SEM image in Fig. 1(c) shows a wire width ~ 10 nm but the theoretical calculations suggest wire width of 8 nm. The difference could be explained by taking into consideration surface depletion effects which leads to narrowing of the wire dimension. The possibility of interaction between the quantum wells does not arise due to the 10 nm thick barrier between them.

### 3. Summary

In the present letter we have used the well-known technique of wet etching to create naturally formed nanowires of controllable widths in an InGaN-based LED heterostructure. The wire dimension is tunable through the etching duration. The tightly confined electron-hole pairs in the narrow essentially defect-free wires tend to form stable excitons compared to those in the quantum well LED heterosturcture. The nano-fabricated LED reveals super-linear response at current densities as high as 400 A/cm$^2$. The light output power vs. bias current density plot shows a small variation with temperature as expected from a predominantly defect-free nanowire LED. The innovative technique of nano-wire formation demonstrated here without requisitioning the use of resource intensive growth methods will make research in the area of quantum confined structures in the technologically important class of III-nitrides to be inexpensive and simple. Various other devices can be fashioned from these naturally selected lateral nanowires like high electron mobility transistors and lasers where material defects need to be minimized.

**Supplementary Material**

The threading dislocations in the material induce anisotropic etching in its vicinity. The screw type dislocations serve as non-radiative recombination centres. The SEM imaging can be used to distinguish the different types of dislocation centres: namely edge, screw and mixed

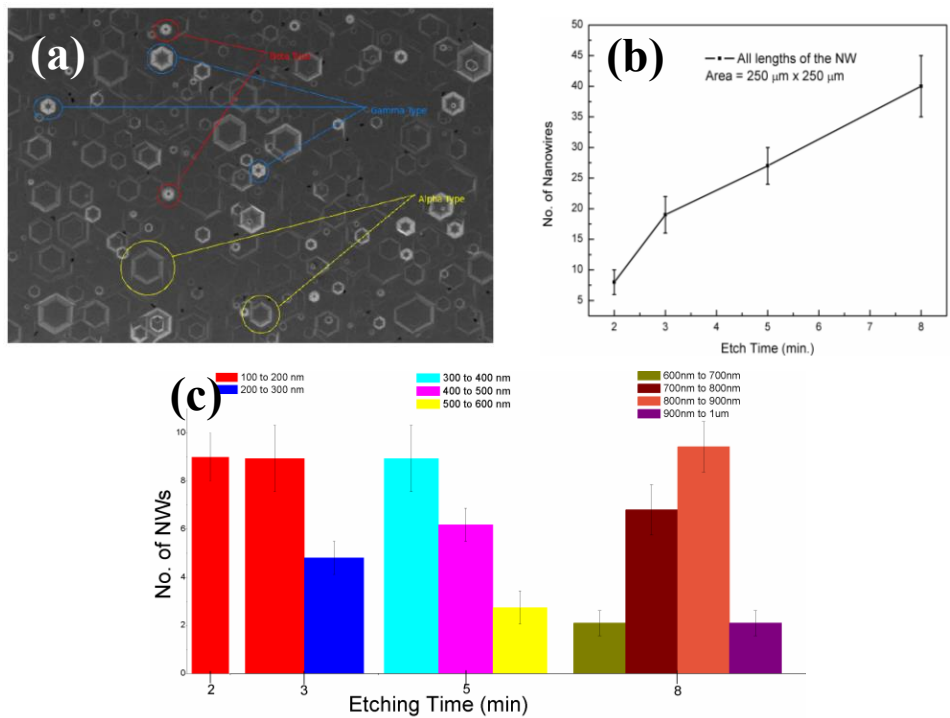

Fig. 1 (a) Scanning Electron Microscope (SEM) image of the GaN surface after etching in 85 % concentrated $H_3PO_4$ showing α, β and γ type of etch pits. The α-type pits are thought to have originated from screw dislocations, β-type from edge dislocations and γ-type from mixed threading dislocations [17]. (b) The plot of the number of nanowires per unit area formed with increasing etching time. (c) The histogram of the number of nanowires of a particular length as a function of the etching time. As the etching is prolonged, the total number of nanowires per unit area increases. It also leads to longer wires.

threading dislocations [17 of main text]. The various types of threading dislocations can be seen in Fig. 1(a). The screw dislocations are a primary reason for the non-radiative losses in light-emitting devices. A statistical survey of the etched surface shows that both the number of quantum wires per unit area and the length of the nanowires increase with increasing etching time, as seen in Fig. 1(b) and (c).

It is observed that the temperature dependent blue shift of the photoluminescence (PL) primary peak energy increases with etching time, as can be seen in Fig. 2. The primary PL peak of the unetched sample has a maximum blue shift of less than 1 meV; that of the 5 min. and 8 min. etched samples are 5 meV and 15 meV respectively. It is also observed that the temperature dependent blue shift manifests at a higher temperature, denoted by an arrow in Fig. 2(b) and (c), for longer etching durations. As the etching duration is prolonged the surface wire density increases since more number of etch pits tend to merge and already formed nanowires become narrower. Thus, the spatial density of localized quantum confined states increase. Another outcome of the prolonged etching is that the lateral wire dimension diminishes precipitating in increased confinement energy. As the excitons thermalize out of the conduction band minimas possibly formed by Indium fluctuation, the increasingly higher energy of the conduction band-edge, brought about by confinement effects, lead to a greater blue shift in the emitted light when the excitons annihilate by radiative recombination. Additionally, the higher confinement energy in narrower wires requires greater thermal energy for the excitons to be able to move out of the potential fluctuation induced states. Hence the observed high temperature shift of the point where the blue shift starts. Figure 3 is the PL spectrum of sample C at 150 K as a function of excitation power density.

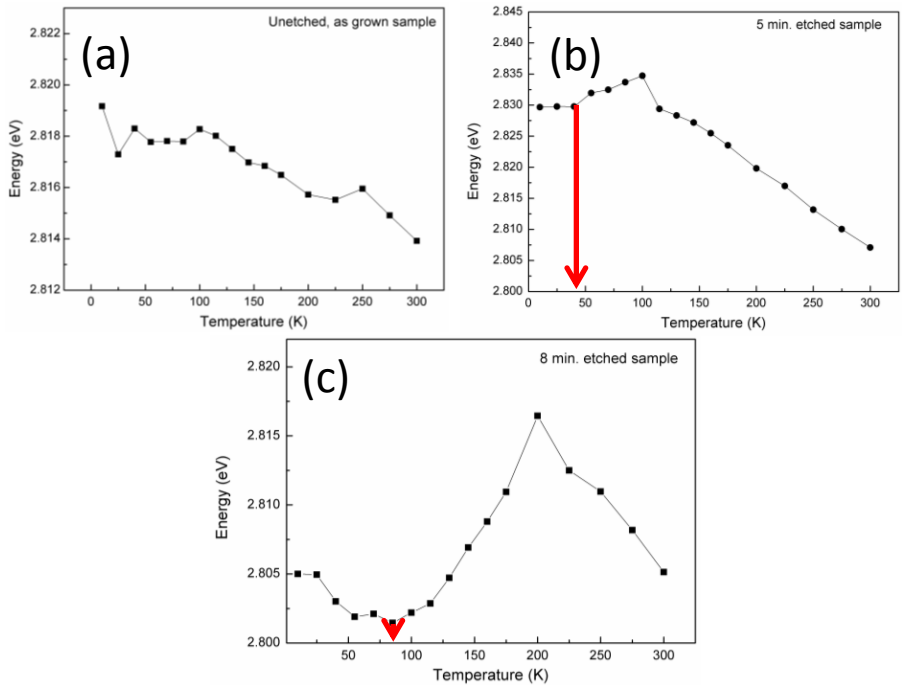

Fig. 2 The PL primary peak energy versus temperature of the (a) unetched, (b) 5 min. etched and (c) 8 min. etched samples. The S-type character of the plot is enhanced with etching time.

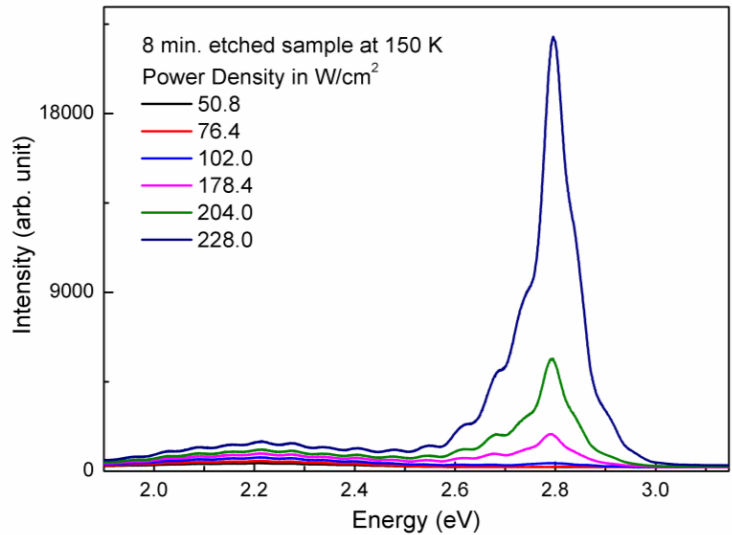

Fig. 3 Power dependent PL at 150 K of sample C. The saturation of the yellow emission band is evident while the blue emission associated with excitonic optical transitions increases exponentially.

We have isolated a single nanowire by mesa definition and dry etching the surrounding material. The isolated quantum wire sample, labelled D is wet etched for 8 minutes similar to sample C. The power dependent photoluminescence was performed on this sample at room temperature, as shown in Fig. 4. It is observed that the line-width is narrow and the primary peak does not saturate with excitation power density.

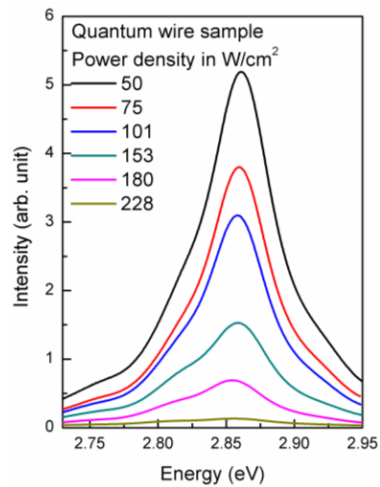

Figure 4 Power dependent photoluminescence spectrum of an isolated quantum wire, sample D.

The non-equilibrium band diagram in Figure 5, with only the conduction band shown, shows that there is a complete attenuation of the electron wave-function in the barrier layers of the quantum well heterostructure. Thus chances of inter-well interaction is negligible. The emission characteristics of a second representative device are given in Fig. 6.

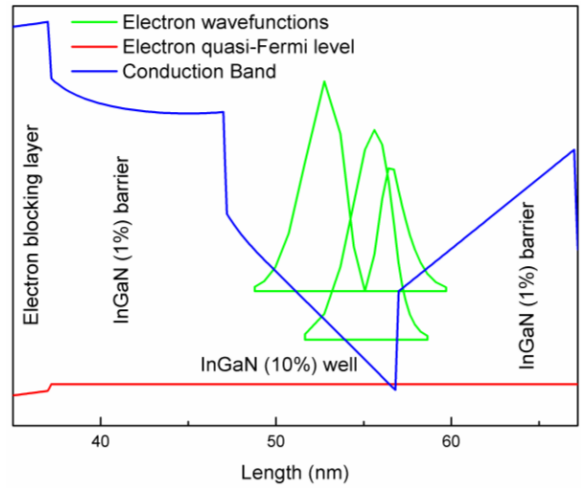

Fig. 5 The non-equilibrium band diagram with the electron bound states penetrating into the barrier.

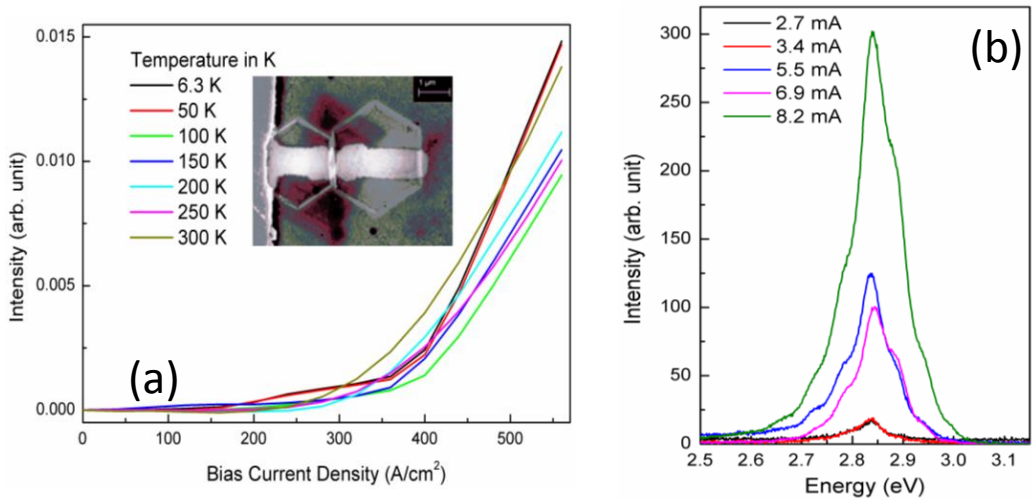

Fig. 6 Emission characteristics of a second representative device (a) Light output power versus bias current density, inset shows the SEM image of the device and (b) electroluminescence spectrum of the nanowire LED device.

The nanowires can be used to further improve the performance of the spin-laser using the Cr-doped GaN as the spin injector [1]. Spin lasers are predicted to have a better performance in terms of lower threshold currents, higher modulation bandwidth and polarization control through spin injection [2-3].